# Great enhancement of strong-field ionization in femtosecond-laser subwavelength-structured fused silica

Shaohua Ye and Min Huang*

State Key Laboratory of Optoelectronic Materials and Technologies and School of Physics, Sun Yat-Sen University, Guangzhou, 510275, China

*Corresponding author: syshm@163.com

**A wavelength-degenerate pump-probe spectroscopy is used to study the ultrafast dynamics of strong-field ionization in femtosecond-laser subwavelength-structured fused silica. The comparative spectra demonstrate that femtosecond-laser subwavelength structuring always give rise to great enhancement for strong-field ionization as well as third-order nonlinear optical effects, which is the direct evidence of great local field enhancement in subwavelength apertures of fs-laser highly-excited surface. In short, the study shows the prominent subwavelength spatial effect of strong-field ionization in femtosecond-laser ablation of dielectrics, which greatly contributes to the well-known "incubation effect".**

In femtosecond (fs) laser ablation, the formation of subwavelength surface structures is a universal phenomenon occurring on various solid materials [1-4]. It is generally accepted that the near-metallic characteristics of irradiated surface for laser-generated abundant free electrons play an important role for the subwavelength structure formation on many materials [5-14]. Such a transient, active near-metallic state is apt to couple with the incident laser fields and induce extraordinary local field enhancement of polarization dependence via the excitation of propagating or localized plasmonic modes [5-14]. For the highly-nonlinear relationship between the laser intensity and the photoionization rate, the local field enhancement would bring about stronger local ionization enhancement, which may result in subwavelength apertures via ultrafast nano-ablation in the ionization "hot spots (lines)" [5,6]. However, to our knowledge, the direct experimental observation of the transient ionization enhancement in the process of fs-laser-induced subwavelength structures is still lacking, which is the key point to verify the proposal mechanisms for the formation of the structures.

On the other hand, previous studies towards fs-laser ionization of solids often focus on duration effect of imposed laser pulses [15-20], and ignore spatial effect of irradiated material. Actually, concerning on the typical subwavelength features of fs-laser-induced structures, it can be deduced that for fs-laser ionization of solids, in particular dielectrics, a strong self-induced ionization spatial localization may appear—for fs-laser ionization of solids, spatial effects should be paid special attention to, which is critical for understanding damage-threshold fluence ($F_{th}$) decreasing in fs-laser multi-pulse ablation known as the "incubation effect" [1]. However, in previous transient detections of fs-laser ionization of solids, spatial effects were not explicitly involved.

In the study, the femtosecond wavelength-degenerate pump-probe transmission spectroscopy (FWPTS) [21] is used to study the ultrafast dynamics of strong-field ionization (SFI) in fs-laser subwavelength-structured fused silica. At the irradiation fluences slightly below $F_{th}$, the effect of fs-laser surface subwavelength structuring can be quantitatively evaluated by comparing the transient transmission spectra between the ablated surface and the non-ablated surface. Such transient detections can provide direct experimental evidence for the previous proposed mechanisms about the extraordinary local field enhancement in subwavelength structures irradiated by near-$F_{th}$ fs-laser pulses.

Here FWPTS was carried out with a 795 nm, 90 fs, 1 kHz Ti:sapphire laser (COHERENT, Legend Elite USP HE+). One can refer to Ref. [20] for more details about the technique and setup of FWPTS. An optical polishing fused silica sheet (KMT, 1 inch diameter, 1 mm thickness, and < 5 Å surface roughness) was used as the sample. Local subwavelength-structured areas on the front surface of the sample were obtained by fixed-point ablation from the pump beam with irradiated fluence slightly higher than $F_{th}$. By controlling the ablation time (the pulse number), a series of ablated craters with varying structured area and degree was achieved. After surface structuring of certain ablation time, onsite FWPTS measurement was carried out to evaluate the effect of surface structuring on fs-laser ionization via comparing with that of an untreated area. For the "incubation effect" [1], $F_{th}$ of a structured surface decreases in comparison with that of an optical polishing surface. Moreover, towards FWPTS the fluence of pump beam should control below $F_{th}$, avoiding secondary damage in the detected process. Therefore, for FWPTS of the structured surface, the fluence of pump beam should be weakened to a suitable range below the new $F_{th}$ (<1) of the structured surface. Resembling Ref. [21], the four representative polarization settings are denoted as "H+H", "H+V", "V+H", and "V+V" according to the pump-probe polarization combinations. Concerning the light-collecting setting [21], a high numerical aperture (NA) of 0.35 ("HNA") and a low NA of 0.12 ("LNA") are utilized alternatively for emphasizing on specific physical factors. Note that the normalized pump fluence $F_n=F/F_{th}$ is used instead of the actual fluence $F$ for quantitatively evaluating the critical ionization characteristic in terms of $F_{th}$.

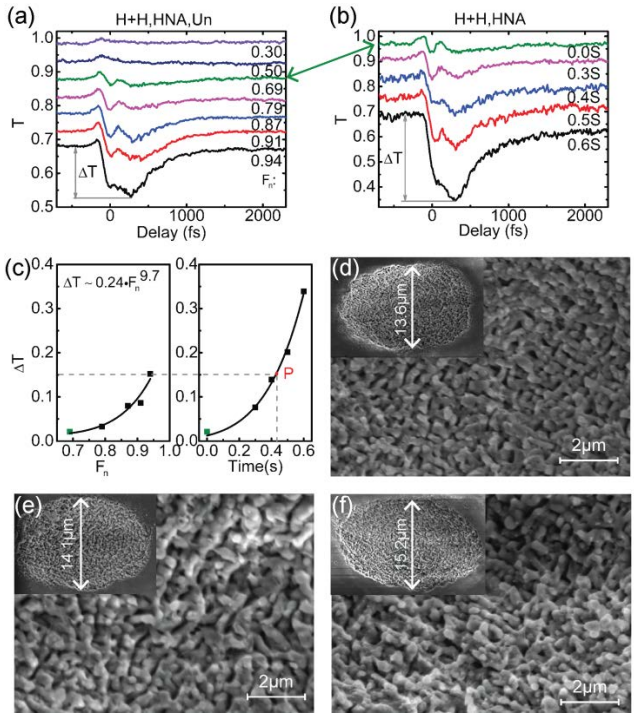

**Fig. 1.** In the H+H and HNA settings FWPTS of fused silica toward (a) an untreated area for a series of increasing $F_n$ approaching 1 (damage threshold), and (b) treated areas for a series of increasing ablation time (the green curve is the initial state without ablation). (c) The evolutions of $\Delta T$ for the curve series in (a) and (b) are displayed on the left and right columns, respectively. SEM images of (d), (e), and (f) show the morphologies of the areas ablated by 0.4, 0.5, and 0.6 s, respectively (the inset shows the whole ablated crater).

In Fig. 1, for the polarization-degenerate setting H+H and the light collecting setting HNA, a comparative FWPTS study of the untreated and treated areas are presented. Towards the untreated area, Fig. 1(a) shows that as $F_n$ increases from 0.30 to 0.94, two recessed regions emerge and become more and more prominent, which are typical spectral features of FWPTS for SFI in fused silica [21]: the two dynamic processes are ascribed to the third-order nonlinear optical effects (TNOEs, occurring at the time zero of pump-probe) and effects of free electron plasma generated by SFI (including initial optical-field ionization and subsequent impact ionization), corresponding to the instantaneous effects of laser optical field and the delayed effects of free electron dynamics, respectively. In the left column of Fig. 1(c), the relationship between $\Delta T$ of the delayed SFI process and $F_n$ exhibits a fast growth trend with a power exponent of 9.7 for the fitting power function, indicating the high nonlinearity of SFI. As a contrast, with the same pump $F_n$ 0.69 avoiding secondary damage, FWPTS of the structured areas prepared with ablation $F_n$ 1.20 are shown in Fig. 1(b) for a series of increasing ablation time of 0.3, 0.4, 0.5 and 0.6 s (as the SEM images shown in Fig. 1(d)-(f), treated areas appear the typical morphology of a fs-laser ablation crater covered by laser-induced subwavelength structures, which present a certain degree of randomness owing to the physical properties of fused silica). Significantly, here with the increasing of ablation time obvious signal enhancement appears for both two dynamic processes. Correspondingly, in the right column of Fig. 1(c), the relationship between $\Delta T$ and ablation time exhibits a fast growth trend, resembling that between $\Delta T$ and $F_n$ in the left column. Note that a much larger $\Delta T$ range for the structured surface can be achieved than that for the untreated surface, and over fifteen-fold growth of $\Delta T$ arises for the ablation time of 0.6 s. These distinct results confirm that the fs-laser subwavelength structuring can significantly enhance SFI as well as TNOEs, and the enhancement effect strongly depends on the ablation degree. Considering the instantaneous respond of TNOEs to the laser field, it is straightforward for us to ascribe the great enhancement of SFI in the structured surface to the great local field enhancement in the subwavelength apertures of fs-laser highly-excited surface [5,6,11].

In view of the similar growth trend of the two curves in Fig. 1(c), a qualitative comparison between the effects of $F_n$ and ablation time on $\Delta T$ (the degree of SFI) can be made. First, a near sevenfold growth of $\Delta T$ is achieved with the increasing of $F_n$ from 0.69 to 0.94. Correspondingly, a similar growth amplitude of $\Delta T$ can be realized with the ablation time about 0.42 s, as marked by the red point P in the fitted curve. Therefore, it can be concluded that the average field enhancement of treated surface arose by subwavelength structuring with the ablation time of 0.42 s is analogous to the direct field enhancement of untreated surface arose by pump $F_n$ increasing from 0.69 to 0.94. Note that it is not achievable for us to get the actual field enhancement value directly from such a comparison, because the effective structured area is increasing with ablation time (the three typical craters in Fig. 1(d)-(f) present the increasing diameters of 13.6, 14.1, and 15.2 μm). Nevertheless, by comparing the change of $\Delta T$ with the change of the structured area, one can see that the relationship between $\Delta T$ and the structured area is nonlinear. Actually, as ablation time increases, not only the structured area will increase, but also the feature size of subwavelength structures will decrease—the scaling-down of laser-induced subwavelength structures will strengthen the local field enhancement greatly [11]. In short, with the increasing of ablation time, the spatial extending of the crater and the spontaneous scaling-down of subwavelength structures both strongly promote the local field enhancement.

In Fig. 2, the polarization setting H+H and the light collecting setting LNA are used for amplifying the signal amplitude ($\Delta T$), in particular that of TNOEs [21]. Pump pulses with $F_n$ of 1.11 and 0.80 are adopted for the ablation and detection processes, respectively. In Fig. 2(a), the obvious first pit in FWPTS of the untreated surface (0 s) indicates that the signal of TNOEs is greatly enhanced by the LNA setting. Then for the treated surface with ablation time increasing from 0.1 to 0.6 s, $\Delta T$ of both two processes is increasing rapidly due to the spatial extending of the crater and the size reduction of formed structures (see the SEM images in Fig. 2(c) and (d)). Eventually, as Fig. 2(b) exhibited the increasing trend becomes saturated, with a largest $\Delta T$ near 0.6. This is because the size and morphology of the ablation crater tends to be stable after employing sufficient laser pulses. Note that in Fig. 2(a), with the increasing of ablation time the TNOE pit is gradually covered by the SFI pit, resembling the spectral features in Fig. 1(a) and (b) for the near-linear intensity dependence of TNOEs and the highly-nonlinear intensity dependence of SFI [21]. The results of Fig. 2 further confirm the great enhancement of SFI and TNOEs on the fs-laser structured surface, which arises from the local field enhancement of highly-excited subwavelength apertures.

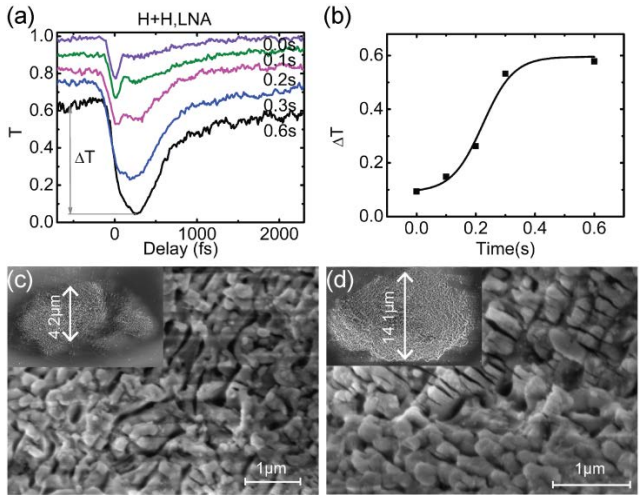

**Fig. 2.** (a) In the H+H and LNA settings FWPTS of an untreated area (0 s) and treated areas with increasing ablation time (0.1, 0.2, 0.3 and 0.6 s). (b) The relationship between ΔT and ablation time. (c) and (d) SEM images of the morphologies of treated areas ablated by 0.2 and 0.3 s, respectively (the inset shows the whole ablated crater).

In Fig. 3, the orthogonal polarization setting H+V and the light-collecting setting HNA are employed to suppress TNOEs and thus get more pure detection of SFI process. In Fig. 3(a), FWPTS of the untreated area present a weaker TNOE pit in contrast with that in Fig. 1(a). In the left column of Fig. 3(c), the fitting function with a power exponent of 6.4 for the relationship between ΔT and $F_n$ also indicates the highly-nonlinear intensity dependence of SFI. In comparison with the untreated case of $F_n$ 0.75 (the navy curve), FWPTS of the structured areas prepared by ablation $F_n$ of 1.20 and ablation time of 0.2, 0.3, 0.6, 1.0, 1.5, and 80 s are shown in Fig. 3(b) (SEM images for the cases of 0.6, 1.5 and 80 s are demonstrated in Fig. 3(d), (e), and (f)). Resembling Fig. 1(b) and 2(a), here with the increasing of ablation time a fast growth of ΔT can also be observed, which originates from the evolution of ablation crater as shown in Fig. 3(d) and (e). As denoted by the red point P in Fig. 3(c), the signal enhancement of untreated surface arose by pump $F_n$ increasing from 0.75 to 0.94 can also be realized via surface structuring with ablation time about 1.50 s. When the ablation time is large enough, the increasing trend of ΔT becomes saturated for reaching the stable state of ablation crater (see the size-change trend of the craters shown in Fig. 3(d)-(f)), like the case in Fig. 2. In general, the trend of SFI enhancement obtained in the H+V polarization setting is similar to that obtained in the H+H polarization setting. However, concerning TNOEs in Fig. 3(b), one can see that after the surface structuring, the TNOE signal is greatly suppressed and almost disappears in the strongly structured areas. It turns out that the surface structuring further weaken the third-order nonlinear interaction between the pump and probe beams in the H+V polarization setting, for the polarization dependence and superficial localization of field enhancement in the fs-laser induced subwavelength structures.

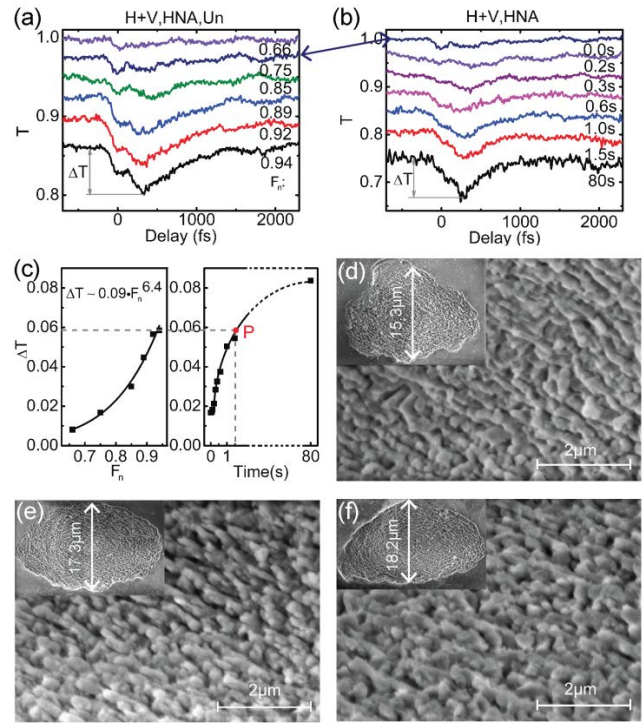

**Fig. 3.** In the H+V and HNA settings FWPTS of (a) an untreated area and (b) treated areas with increasing ablation time (the blue curve is the initial state without ablation). (c) The evolutions of ΔT for the curve series in (a) and (b) are displayed on the left and right columns, respectively. SEM images of (d)-(f) show the morphologies of the areas ablated by 0.6, 1.5, and 80 s (the inset shows the whole ablated crater).

Above experiments focus on the relationship between SFI enhancement and ablation time for a constant $F_n$. In the following, the relationship between SFI enhancement and $F_n$ for a stable crater (Fig. 4(a)) ablated with $F_n$ 1.10 and ablation time 5 s is probed into. Towards the crater, FWPTS for four typical polarization combinations (H+H, H+V, V+V, and V+H) are shown in Fig. 4(c)-(f) with maximal pump $F_n$ restricted to 0.84, avoiding secondary damage. In terms of the spectra of untreated surface in Fig. 4(b), one can clearly see that after the surface structuring, for all polarization cases SFI signal is greatly enhanced in the low and moderate $F_n$ range; then as $F_n$ approaches 1, the signal gradually stabilizes—the trend is different from the untreated surface.

To get a more direct comparison for the four polarization settings, the curves for $F_n$ 0.63 in Fig. 4(b)-(f) are exhibited together in Fig. 4(g). On the whole, comparing with the curve for the untreated surface, all the four curves for the structured surface present a similar, prominent SFI enhancement (the drift of initial descent stage comes from different delays introduced by half-wave plates for polarization settings and different responses of TNOEs in four cases). Concerning ΔT, the H+H and H+V cases are larger than the V+H and V+V cases, demonstrating the effect of polarization direction of pump beam on SFI enhancement for the subwavelength structures provided with a certain degree of directionality. The result is easy to understand in view of the polarization dependence of field enhancement in highly-excited subwavelength structures [5]. Whereas, the effect of polarization directions of probe beam is far weaker—the SFI signal coming from the inverse bremsstrahlung absorption of probe beam is not sensitive to the polarization of probe beam. Further, for all the spectra in Fig. 4(b)-(f) the evolution of $\Delta T$ as a function of $F_n$ are exhibited together in Fig. 4(h). In the $F_n$ range from 0.49 to 0.84, the four curves of the treated surface degenerate to two groups associated with the polarization setting of pump beam, further confirming the effect of polarization direction of pump beam on SFI enhancement. Moreover, interestingly, with the increase of $F_n$, $\Delta T$ grows as functions of power and linearity for the untreated and treated cases, respectively. The result turns out that for SFI of a sufficiently-ablated crater with stable morphology, the nonlinear intensity dependence will return to the linear intensity dependence, which indicates the ending of ablation process. Specifically, towards the H+H cases, the ratio (red curve) of the $F_n$-$\Delta T$ fitted curves for the treated and untreated areas are shown in Fig. 4(i), which clearly exhibits the evolution of signal enhancement as a function of $F_n$. Significantly, a maximum value near 18 can be obtained with $F_n$ 0.57. That is, for the structured surface irradiated by pump beam with $F_n$ 0.57, the maximum field enhancement effect can be achieved. In other words, for a stable ablation crater, the field enhancement reaches the maximum at the sub-damage-threshold fluence. The result is consistent with the theory of the excitation of quasistatic surface plasmons in deep-subwavelength apertures [11], which appears in the electrostatic regime with extraordinary electrostatic field enhancement.

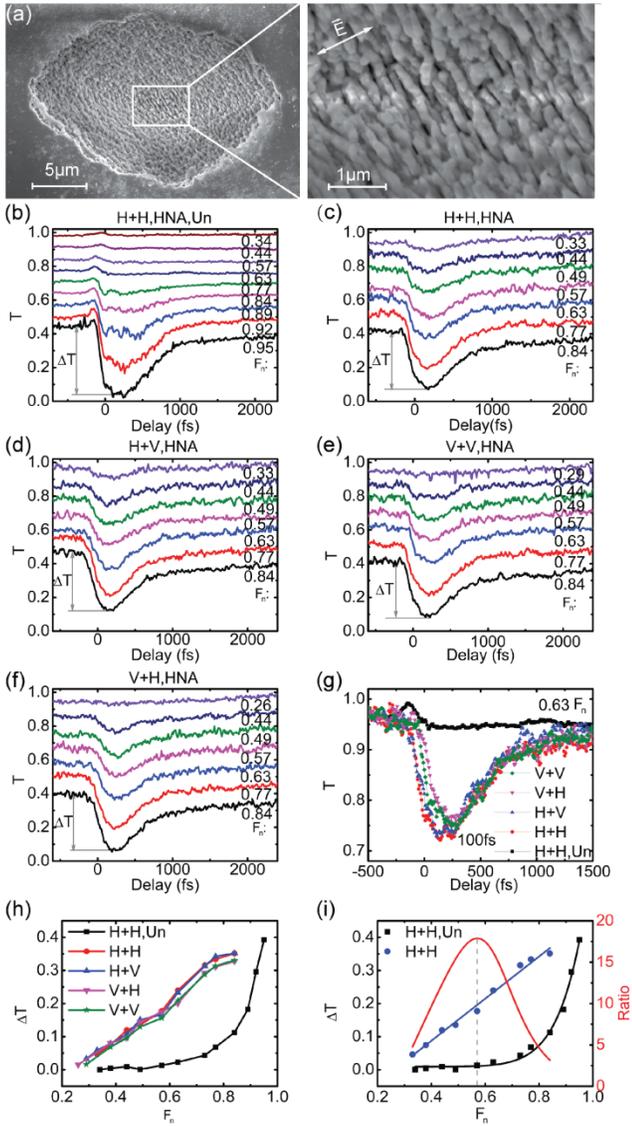

**Fig. 4.** (a) SEM images of the morphology of a stable crater ablated with $F_n$ 1.10 and ablation time 5 s. FWPTS of (b) an untreated area and treated areas for four typical polarization combinations of (c) H+H, (d) H+V, (e) V+V, and (f) V+H. (g) The curves for $F_n$ 0.63 in (b)-(f) are exhibited together. (h) The evolution of $\Delta T$ as a function of $F_n$ for all the spectra in (b)-(f). (i) Towards the H+H cases in (h), the ratio of the $F_n$-$\Delta T$ fitted curves for the treated and untreated areas.

In brief, the study focuses on the spatial effect of SFI in fs-laser subwavelength-structured fused silica. FWPTS confirm that fs-laser subwavelength structuring always leads to great enhancement for SFI as well as TNOEs, which serves as direct evidence for great local field enhancement in fs-laser highly-excited subwavelength apertures and provides new insights into the origin of "incubation effect" in fs-laser ablation.

**Funding.** National Natural Science Foundation of China (NSFC) (11274400); Pearl River S&T Nova Program of Guangzhou (201506010059).

**REFERENCES**

1. J. Bonse, S. Baudach, J. Krüger, W. Kautek, and M. Lenzner, Appl. Phys. A **74**, 19 (2002).
2. A. Borowiec and H. Haugen, Appl. Phys. Lett. **82**, 4462 (2003).
3. J. Bonse, A. Munz, and H. Sturm, J. Appl. Phys. **97**, 013538 (2005).
4. T. Jia, H. Chen, M. Huang, F. Zhao, J. Qiu, R. Li, X. Xu, X. He, J. Zhang, and H. Kuroda, Phys. Rev. B **72**, 125429 (2005).
5. M. Huang, F. Zhao, Y. Cheng, N. Xu, and Z. Xu, Phys. Rev. B **79**, 125436 (2009).
6. M. Huang, F. Zhao, Y. Cheng, N. Xu, and Z. Xu, ACS Nano **3**, 4062 (2009).
7. J. Bonse, A. Rosenfeld, and J. Krüger, J. Appl. Phys. **106**, 104910 (2009).
8. F. Garrelie, J. P. Colombier, F. Pigeon, S. Tonchev, N. Faure, M. Bounhalli, S. Reynaud, and O. Parriaux, Opt. Express **19**, 9035 (2011).
9. J. Bonse, J. Krüger, S. Höhm, and A. Rosenfeld, J. Laser Appl. **24**, 042006 (2012).
10. M. Huang, F. Zhao, Y. Cheng, and Z. Xu, Ann. Phys. (Berlin) **525**, 74 (2013).
11. M. Huang and Z. Xu, Laser Photonics Rev. **8**, 633 (2014).
12. Y. Liao, J. Ni, L. Qiao, M. Huang, Y. Bellouard, K. Sugioka, and Y. Cheng, Optica **2**, 329 (2015).
13. J. Song, W. Tao, H. Song, M. Gong, G. Ma, Y. Dai, Q. Zhao, J. Qiu, Appl. Phys. A **122**, 341 (2016).
14. K. Zhou, X. Jia, T. Jia, K. Cheng, K. Cao, S. Zhang, D. Feng, and Z. Sun, J. Appl. Phys. **121**, 104301 (2017).
15. B. C. Stuart, M. D. Feit, S. Herman, A. M. Rubenchik, B. W. Shore, and M. D. Perry, Phys. Rev. B **53**, 1749 (1996).
16. M. Lenzner, J. Krüger, S. Sartania, Z. Cheng, C. Spielmann, G. Mourou, W. Kautek, and F. Krausz, Phys. Rev. Lett. **80**, 4076 (1998).
17. A. C. Tien, S. Backus, H. Kapteyn, M. Murnane, and G. Mourou, Phys. Rev. Lett. **82**, 3883 (1999).
18. A. Kaiser, B. Rethfeld, M. Vicanek, and G. Simon, Phys. Rev. B **61**, 11437 (2000).
19. G. M. Petrov and J. Davis, J. Phys. B **41**, 025601 (2008).
20. B. Chimier, O. Utéza, N. Sanner, M. Sentis, T. Itina, P. Lassonde, F. Légaré, F. Vidal, and J. C. Kieffer, Phys. Rev. B **84**, 2669 (2011).
21. S. Ye and M. Huang, arXiv: 1803.02566 (2018).